\documentclass[preprint,12pt]{elsarticle}
\usepackage{graphicx}
\usepackage{amssymb}
\usepackage{amsmath}

\begin{document}
\title{Diffusion in active magnetic colloids}
%\author{R.Taukulis, A.Cebers}
%\affiliation{University Paris Sud,France}
%\author{A.C\={e}bers}
%\affiliation{University of Latvia, Ze\c{l}\c{l}u-8, R\={\i}ga, %LV-1002, Latvia}

\date{\today}

\author{R.Taukulis, A.Cebers}

\address{University of Latvia, Ze\c{l}\c{l}u-8, R\={\i}ga, LV-1002, Latvia}

\begin{keyword}
Active systems Magnetotactic bacteria Circle swimmer  Diffusion
%% keywords here, in the form: keyword \sep keyword

%% MSC codes here, in the form: \MSC code \sep code
%% or \MSC[2008] code \sep code (2000 is the default)

\end{keyword}
\begin{abstract}
Properties of active colloids of circle swimmers are reviewed. As an particular example of active magnetic colloids the magnetotactic bacteria under the action of a rotating magnetic field is considered. The relation for a diffusion coefficient due to the random switching of the direction of rotation of their rotary motors is derived on the basis of the master equation. The obtained relation is confirmed by the direct numerical simulation of random trajectory of a magnetotactic bacterium under the action of the Poisson type internal noise due to the random switching of rotary motors. The results obtained are in qualitative and quantitative agreement with the available experimental results and allows one to determine the characteristic time between the switching events of a rotary motor of the bacterium.
\end{abstract}

%\pacs{47.57.J-,47.20.Ky,83.80.Gv}

\maketitle
Properties of random trajectories of so-called circle swimmers have been studied by various groups of researchers \cite{1,2,3,4,5}. Various kinds of these swimmers and their properties have been investigated. Effective diffusion coefficients of circle swimmers subjected to thermal noise are calculated in \cite{1,2}. Circle swimmers with an asymmetric L-shape self-propelling due to the self-phoresis at illumination by light were synthesized and studied in \cite{4}. Circular dynamics of slightly bent self-propelling rods switching their direction of circulation due to thermal fluctuations are studied in \cite{5}.

A specific example of circle swimmers is magnetotactic bacteria, which in a rotating magnetic field in the absence of noise  move along circular trajectories, if the frequency of the rotating field is less than a critical value \cite{6}. A random walk of the centers of circular trajectory is observed due to random switching of the direction of rotation of their flagella \cite{6,7}. Here the properties of this random walk are studied on the basis of the master equation \cite{5,8} for the probability density distribution function $f(\vec{x},\vec{n},t)$ of position of the bacterium $\vec{x}$ and direction of its magnetic moment $\vec{n}$. Bacteria moves in $\vec{n}$ direction or opposite to it. It is illustrated that this method may be applied also to circle swimmers of other types.
\section{Master equation}
Let us consider magnetotactic bacteria which move in the direction of their magnetic moment $\vec{n}$ or opposite to it due to random switching of the direction of rotation of a rotary motor.
Since in the synchronous regime the magnetic moment rotates synchronously with the field, so does the velocity of the bacterium. The master equation accounting for a random switching rate $\lambda$ of the direction of a rotary motor rotation reads
\begin{equation}
\frac{\partial f(\vec{x},\vec{n},t)}{\partial t}=-\frac{\partial (v\vec{n}f(\vec{x},\vec{n},t))}{\partial \vec{x}}-\frac{\partial (\vec{\omega}\times \vec{v}f(\vec{x},\vec{n},t))}{\partial \vec{v}}-\lambda f(\vec{x},\vec{n},t)+\lambda f(\vec{x},-\vec{n},t)~.
\label{Eq:1}
\end{equation}
Taking into account that $\vec{v}\times \partial / \partial \vec{v}=\partial /\partial \vartheta$, where $\vec{v}=v\vec{n}=v(\cos{\vartheta},\sin{\vartheta})$, Eq. (\ref{Eq:1}) may be rewritten for the two-dimensional case in a more simple way as follows
\begin{equation}
\frac{\partial f(\vec{x},\vec{n},t)}{\partial t}=-\frac{\partial (v\vec{n}f(\vec{x},\vec{n},t))}{\partial \vec{x}}-\omega\frac{\partial (f(\vec{x},\vec{n},t))}{\partial \vartheta}-\lambda f(\vec{x},\vec{n},t)+\lambda f(\vec{x},-\vec{n},t)~.
\label{Eq:1a}
\end{equation}

There are several possible ways to analyse the properties of random trajectories of these circle swimmers. A simple approach is based on the derivation from (\ref{Eq:1a}) of a closed set of equations for several moments characterizing their distribution $<\vec{x}^{2}>,<\vec{x}\vec{n}>,<\vec{x}d\vec{n}/d\vartheta>$. It reads
\begin{eqnarray}
\frac{d<\vec{x}^{2}>}{dt}=2v<\vec{x}\vec{n}>~, \\ \nonumber
\frac{d<\vec{x}\vec{n}>}{dt}=v+\omega<\vec{x}\frac{d\vec{n}}{d\vartheta}>-2\lambda<\vec{x}\vec{n}>~, \\ \nonumber
\frac{d}{dt}<\vec{x}\frac{d\vec{n}}{d\vartheta}>=-\omega<\vec{x}\vec{n}>-2\lambda<\vec{x}\frac{d\vec{n}}{d\vartheta}>~.
\label{Eq:3}
\end{eqnarray}
In order to derive a set of equations (3) the condition of normalization $\int f(\vec{x},\vec{n},t)d\vec{x}d^{2}\vec{n}=1$ and the relation $d^{2}\vec{n}/d\vartheta^{2}=-\vec{n}$ are used. Since the set (3) is closed it is possible to obtain an analytical solution for $<\vec{x}^{2}>$, which characterizes the random process of the particle diffusion.

The stationary solution for the moments $<\vec{x}\vec{n}>,<\vec{x}\frac{d\vec{n}}{d\vartheta}>$ reads
\begin{equation}
<\vec{x}\vec{n}>=\frac{2v \lambda}{(2\lambda)^{2}+\omega^{2}}~,
\label{Eq:4}
\end{equation}
\begin{equation}
<\vec{x}\frac{d\vec{n}}{d\vartheta}>=-\frac{v\omega}{(2\lambda)^{2}+\omega^{2}}~.
\label{Eq:5}
\end{equation}
Relations (\ref{Eq:4},\ref{Eq:5}), as a result, give
\begin{equation}
\frac{d<\vec{x}^{2}>}{dt}=4\frac{v^{2}\lambda}{(2\lambda)^{2}+\omega^{2}}~.
\label{Eq:6}
\end{equation}
Taking into account 2D character of particle motion, absence of cross-correlation $<xy>$ in the stationary case and $<x^{2}>=<y^{2}>$ due to the symmetry, for the diffusion coefficient we have
\begin{equation}
D=\frac{v^{2}\lambda}{(2\lambda)^{2}+\omega^{2}}~.
\label{Eq:7}
\end{equation}
In the limit $\omega\rightarrow 0$ we have
\begin{equation}
D=\frac{v^{2}}{4\lambda}~.
\label{Eq:8}
\end{equation}
Introducing as a characteristic time of tumbling $\tau_{s}=\lambda^{-1}$ relation (\ref{Eq:8}) reads
\begin{equation}
D=\frac{v^{2}\tau_{s}}{4}~.
\label{Eq:9}
\end{equation}
It coincides with the expression for the diffusion coefficient of tumbling bacterium \cite{9} with $<\cos{\delta}>=-1$, where $\delta$ is angle by which the bacterium changes direction of motion after a tumbling event. Since in our case direction of the bacterium motion changes to the opposite we have $<\cos{\delta}>=-1$. In the limit of high frequency for the diffusion coefficient we have $D=r_{0}^{2}/\tau_{s}$, where $r_{0}$ is the radius of a circular trajectory, which coincides with the expression of the diffusion coefficient in the elementary theory of Brownian motion \cite{10}.
\section{Circle swimmer and thermal noise}
We should remark that besides the diffusion due to random switching of the rotary motor of the bacterium there is also thermal noise due to fluctuating forces and torques. This problem is considered in \cite{1,2}. In \cite{1,2} calculation is carried out by using time-correlation functions. We will illustrate further that the diffusion coefficients of torqued swimmers may be calculated in a similar fashion as above on the basis of the Fokker-Planck equation for the joint distribution function of the position and orientation of the swimmer $f(\vec{r},\vec{n},t)$:
\begin{equation}
\frac{\partial f}{\partial t}=-\frac{\partial(v\vec{n}f)}{\partial \vec{x}}-\vec{\omega}\vec{K}_{\vec{n}}f+D_{R}\vec{K}^{2}_{\vec{n}}f
+D\frac{\partial ^{2} f}{\partial \vec{x}^{2}}
\label{Eq:10}
\end{equation}
where $\vec{K}_{\vec{n}}=\vec{n}\times \partial /\partial \vec{n}$ is the operator of infinitesimal rotations and $D$ and $D_{R}$ are translational and rotational diffusion coefficients respectively. In 3D case the following closed set of equations for the moments may be obtained
\begin{eqnarray}
\frac{d<\vec{x}^{2}>}{dt}=2v<\vec{x}\vec{n}>+6D~,\\ \nonumber
\frac{<x^{2}_{z}>}{dt}=2D+2v<x_{z}n_{z}>~,\\ \nonumber
\frac{d<\vec{x}\vec{n}>}{dt}=v+\omega<(\vec{n}\times\vec{x})_{z}>
-2D_{R}<\vec{x}\vec{n}>~, \\ \nonumber
\frac{d<x_{z}n_{z}>}{dt}=v<n_{z}^{2}>-2D_{R}<x_{z}n_{z}>~,\\ \nonumber
\frac{d<n_{z}^{2}>}{dt}=2D_{R}-6D_{R}<n^{2}_{z}>~,\\ \nonumber
\frac{d<(\vec{n}\times\vec{x})_{z}>}{dt}=-2D_{R}<(\vec{n}\times\vec{x})_{z}>
-\omega<\vec{x}\vec{n}>+\omega<x_{z}n_{z}>~.
\end{eqnarray}
In the stationary case we have
\begin{eqnarray}
<n^{2}_{z}>=\frac{1}{3}~,\\ \nonumber
<n_{z}x_{z}>=\frac{v}{6D_{R}}~,\\ \nonumber
<\vec{x}\vec{n}>=\frac{v(1+(\omega/(2D_{R})^{2}/3}{\omega^{2}/(2D_{R}}+2D_{R}~,\\ \nonumber
\frac{d<x^{2}_{z}>}{dt}=2D+\frac{v^{2}}{3D_{R}}~,\\ \nonumber
\frac{d<\vec{x}_{\perp}^{2}>}{dt}=4D+\frac{2v^{2}}{3D_{R}(1+(\omega/(2D_{R}))^{2}}~.
\end{eqnarray}
This gives, in agreement with \cite{2},
\begin{equation}
D_{\parallel}=D+\frac{v^{2}}{6D_{R}}
\label{Eq:11}
\end{equation}
and
\begin{equation}
D_{\perp}=D+\frac{v^{2}}{6D_{R}(1+(\omega/(2D_{R}))^{2}}~.
\label{Eq:12}
\end{equation}
Similar  to (14) relation in 2D case, up to our knowledge, was for the first time obtained in \cite{1}. 

We see that the frequency dependence of the diffusion coefficient is similar to the one for the magnetotactic bacterium under the action of the Poisson like noise. The role of the switching rate $\lambda$ is determined in this case by the rotary diffusion coefficient of the swimmer $D_{R}$. Taking for the rotational drag coefficient of the bacterium the value $2.4\cdot 10^{-12}~erg.s$ \cite{6} we may estimate the characteristic switching rate due to the rotational diffusion as $0.017 ~s^{-1}$, which is much less than expected due to the switching of the rotary motors of a bacterium. This shows that random trajectories observed in \cite{6} of bacteria in a rotating field are due to the random switching of rotary motors and not due to the random thermal noise. Another evidence for this comes from direct comparison of the trajectory observed in the experiment (see Fig.11 in \cite{6}) and trajectory obtained numerically as described in \cite{7} and shown in Fig.1. Parameters used in generating the trajectory are similar to those of the experiment. Trajectory length is $200~s$, $f=0.5~\mathrm{Hz}$. From the experiment estimated radius of rotation is $27~\mathrm{pixels}=4.8~\mathrm{\mu m}$ and diffusion coefficient $D=7~\mathrm{\mu m^2~s^{-1}}$ which give $v=15.1~\mathrm{\mu m~s^{-1}}$ and, according to (\ref{Eq:7}), $\lambda=0.32~\mathrm{Hz}$.

Concerning the relation (\ref{Eq:7}) we may point out that for given frequency of rotating field $\omega$ the diffusion coefficient has a maximum at definite switching rate $\lambda=\omega/2$. Thus a bacterium, which is in unfavorable conditions and is confined to a definite area by a rotating field has a possibility to increase its diffusion coefficient for seeking more favorable conditions by adopting the switching rate of the rotary motor. It may be worth to notice that optimal switching rate given by the relation $\lambda=\omega/2$ in the case of a rotating field with frequency $f=0.5~Hz$ equals to $1.6~s^{-1}$ - a value close to the characteristic switching and tumbling events of bacteria \cite{11}.
It is interesting to remark that similar mechanism of the enhancement of the diffusion coefficient of marine bacteria was considered in \cite{12} where it was illustrated that time modulation of the swimming velocity may increase the diffusion coefficient by an order of magnitude. Maximum of the diffusion coefficient for the centers of spiral waves in dependence on the correlation time of random noise was found in \cite{13}.
\section{Transitory regime}
The set of equations (3) allows us to consider the transitory regime of establishment of the Brownian diffusion. Seeking the solution exponential in time $\exp{(\sigma t)}$ of linear homogeneous set of equations
\begin{eqnarray}
\frac{d<\vec{x}\vec{n}>}{dt}=v+\omega<\vec{x}\frac{d\vec{n}}{d\vartheta}>-2\lambda<\vec{x}\vec{n}>~, \\ \nonumber
\frac{d}{dt}<\vec{x}\frac{d\vec{n}}{d\vartheta}>=-\omega<\vec{x}\vec{n}>-2\lambda<\vec{x}\frac{d\vec{n}}{d\vartheta}>
\end{eqnarray}
we obtain $\sigma_{\pm}=-2\lambda \pm i\omega$. As a result the solution for these moments satisfying zero initial conditions reads
\begin{equation}
<\vec{x}\vec{n}>(t)=\frac{2v\lambda}{(2\lambda)^{2}+\omega^{2}}+\frac{v}{2\sigma_{+}}\exp{(\sigma_{+}t)}+\frac{v}{2\sigma_{-}}\exp{(\sigma_{-}t)}
\end{equation}
and
\begin{equation}
<\vec{x}\frac{d\vec{n}}{d\vartheta}>(t)=-\frac{v\omega}{(2\lambda)^{2}+\omega^{2}}+i\frac{v}{2\sigma_{+}}\exp{(\sigma_{+}t)}-i\frac{v}{2\sigma_{-}}\exp{(\sigma_{-}t)}~.
\end{equation}
By integration we find the mean square displacement 
\begin{equation}
<\vec{x}^{2}>=\frac{4v^{2}\lambda}{(2\lambda)^{2}+\omega^{2}}t+\frac{v^{2}}{\sigma_{+}^{2}}(\exp{(\sigma_{+}t)}-1)+\frac{v^{2}}{\sigma_{-}^{2}}(\exp{(\sigma_{-}t)}-1)~.
\label{Eq:12a}
\end{equation}
At $t\rightarrow \infty$ for the mean square displacement we
have 
\begin{equation}
<\vec{x}^{2}>=\frac{4v^{2}\lambda}{(2\lambda)^{2}+\omega^{2}}t-v^{2}\frac{(2\lambda)^{2}-\omega^{2}}{((2\lambda)^{2}+\omega^{2})^{2}}~.
\end{equation}
At small times the motion of the bacterium is ballistic $<\vec{x}^{2}>\simeq v^{2}t^{2}$.
\section{Numerical simulation algorithm}
In the synchronous regime tangent vector $\vec{n}=(\cos{(\vartheta)},\sin{(\vartheta)})$ to the trajectory, which we choose along the magnetic moment of a bacterium, rotates with the angular velocity of the field:
\begin{equation}
\frac{d\vec{n}}{dt}=\vec{\omega}\times \vec{n}=\omega\vec{b}~,
\end{equation}
where $\vec{b}=(-\sin{(\vartheta)},\cos{(\vartheta)})$. Since, according to the Frenet equation,
\begin{equation}
\frac{d\vec{n}}{dl}=k\vec{b}
\end{equation}
and $dl/dt=\pm v$ (the bacterium moves along its magnetic moment or opposite to it) we have $\omega\vec{b}=\pm kv\vec{b}$ and thus  $k=\pm \omega/v$. It means that at constant rotary motor rotation direction the trajectory is an arc of circle with radius $v/\omega$. The radius vector of the curvature center of the trajectory is $\vec{R}=\vec{r}+k^{-1}\vec{b}$. It means that the position of the curvature center jumps by $\Delta\vec{R}=\pm 2v\vec{b}/\omega$ at the switching of the rotary motor. For numerical implementation of the algorithm it is useful to introduce complex notations $\tilde{\vec{n}}=\exp{(it)};\tilde{\vec{b}}=i\tilde{\vec{n}};\tilde{\vec{R}}=\vec{X}+i\vec{Y}$ and $\tilde{\vec{r}}=\vec{x}+i\vec{y}$. In dimensionless form the equations for a random trajectory of the bacterium read ($\tau=\omega t;(\bar{\vec{r}},\bar{\vec{R}})=\omega/v(\tilde{\vec{r}},\tilde{\vec{R}})$)
\begin{equation}
\frac{d\bar{\vec{r}}}{dt}=\eta\tilde{\vec{n}};~\tilde{\vec{n}}=\exp{(i\tau)};\frac{d\bar{\vec{R}}}{dt}=\frac{d\eta}{dt}\tilde{\vec{b}}~.
\end{equation}
For numerical simulation of the trajectory a set of switching times $\tau_{i}~(i=1,...N)$ distributed according to the Poisson law $p_{i}=\exp{(-\lambda\tau_{i})}$ are introduced. At time moments $T_{i}=\sum^{i-1}_{j=1} \tau_{j}+\tau_{i}$ the random variable $\eta$ equal to $\pm 1$ changes the sign.

The generated trajectory is used to obtain diffusion coefficient $D$ as a function of the switching rate $\lambda$. Two examples of computed mean square displacement as a function of time difference is shown in Fig.2. It is evident that the relationship is not linear at small times. Slope of the best-fit line at large times is an estimate of $4D$. The estimated diffusion coefficient for a range of $\lambda$ values is shown in Fig.3. All trajectories used consist of $10^6$ switches of swimming direction. The results are in very good agreement with (\ref{Eq:7}).

In \cite{4}, when considering sperm trajectory curvature fluctuations
$k(t)=k_{0}+\xi_{k}(t)$, it is found that they introduce effective diffusion coefficient given by ($k_{0}=1/r_{0}$)
\begin{equation}
D\simeq\frac{1}{4}(vr_{0})^{2}S_{k}(\omega_{0})~,
\label{Eq:20}
\end{equation}
where $S_{k}(\omega_{0})$ is a power spectrum of the curvature fluctuations, found at the circulation frequency $\omega_{0}=v/r_{0}$. In our case of a synchronous regime of motion of bacteria in a rotating field $\omega_{0}$ is equal to the frequency of the rotating field $\omega$. The relation (\ref{Eq:20}) is in agreement with the numerical simulation results. Taking into account that for the Poisson process the probability of even number of switching events in time $t$ is $\cosh{(\lambda t)}\exp{(-\lambda t)}$ and the probability of odd number of switching events is $\sinh{(\lambda t)}\exp{(-\lambda t)}$
for the curvature time correlation function we have $<k(t)k(0)>=r^{-2}_{0}\exp{(-2\lambda t)}$ and the power spectrum of curvature fluctuations is
\begin{equation}
S_{k}(\omega_{0})=\frac{1}{r^{2}_{0}}\frac{4\lambda}{4\lambda^{2}+\omega_{0}^{2}}~.
\label{Eq:21}
\end{equation}
Relations (\ref{Eq:20}) and (\ref{Eq:21}) give the relation (\ref{Eq:7}). 
It should be noticed that according to \cite{4} the relation (\ref{Eq:21}) has a more broad context than a particular example considered here and includes the noise in chemotactic signalling system and other.
\section{Conclusions}
Analysis of the random trajectories of the circle swimmers - the magnetotactic bacteria shows that experimentally observed random trajectories of bacteria are caused by the internal noise of rotary motors and are not due to the thermal fluctuations. Derived on the basis of the master equation relation for the diffusion coefficient is in an excellent agreement with the results of numerical simulations and allows one to determine the characteristic rate of the random switching events of rotary motors from the available experimental data.
\section{Acknowledgements}
Authors acknowledge funding from the European Union though the FP7 project Bio2MaN4MRI (grant No.245542). AC is thankful to organizers of the 2nd French-Brazilian meeting on Nanoscience,Nanotechnology and Nanobiotechnology (December 2012, Brazil).

\begin{figure}
    \includegraphics[width=1.0\columnwidth]{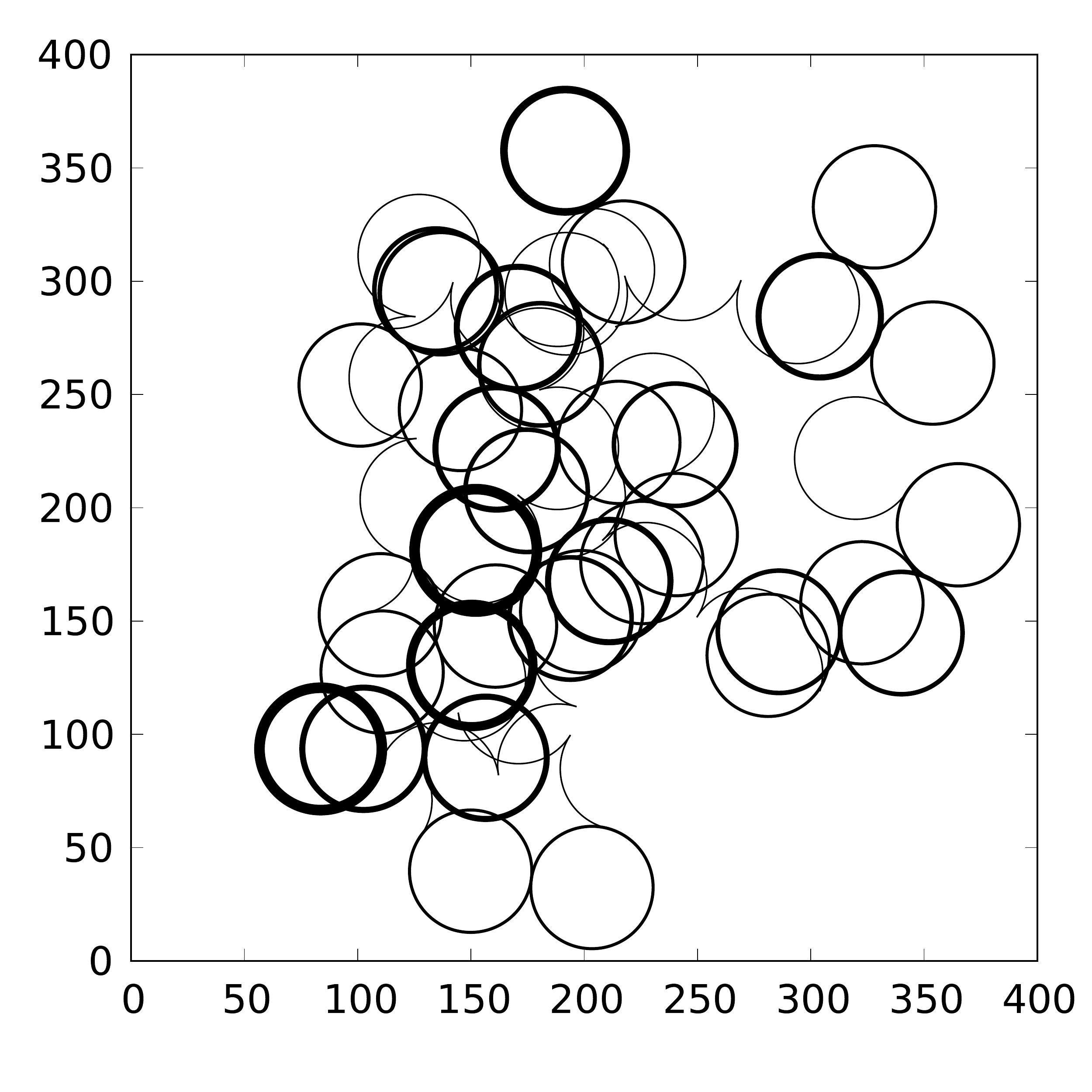}
    \caption{Generated random trajectory of the bacterium. Line thickness is proportional to number of full turns before swithcing of the swimming direction. Coordinates are shown in pixels. 18 pixels = $3.2~\mathrm{\mu m}$. See Fig.11 in \cite{6} for a qualitative and quantitative comparison.
    }
    \label{Fig1}
\end{figure}

\begin{figure}
    \includegraphics[width=1.0\columnwidth]{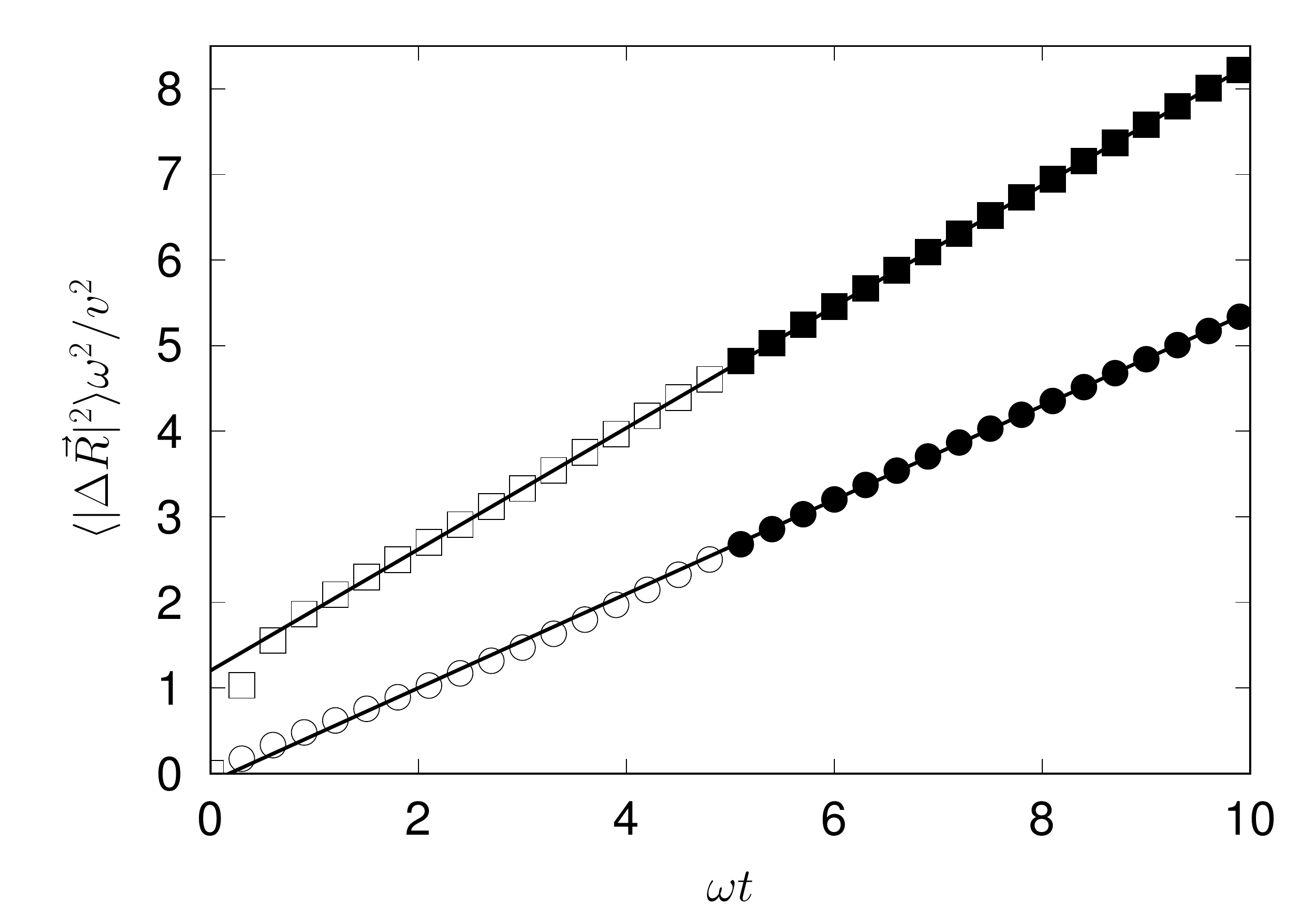}
    \caption{Mean square displacement as a function of time difference. Random trajectories with dimensionless switching rates $\lambda=0.15$ (circles) and $\lambda=1.2$ (squares) are used. For clarity only every sixth data point is shown. Black markers show data used in linear regression.
    }
    \label{Fig2}
\end{figure}

\begin{figure}
    \includegraphics[width=1.0\columnwidth]{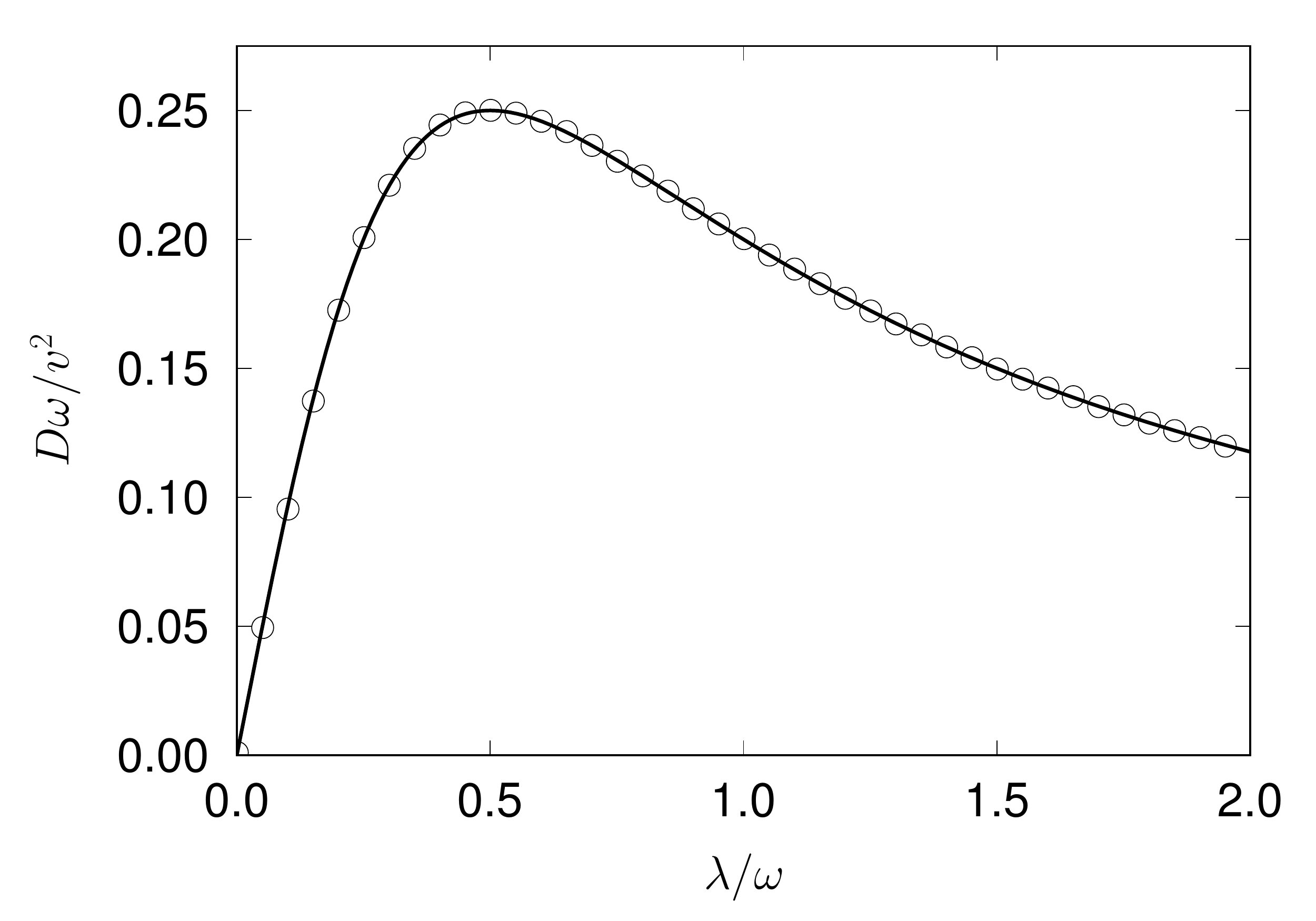}
    \caption{Computed diffusion coefficient (circles) compared to the relation (\ref{Eq:7}).
    }
    \label{Fig3}
\end{figure}


\begin{thebibliography}{15}
\bibitem{1} S.Ebbens, R.A.L.Jones, A.J.Ryan, R.Golestanian, and  J.R.Howse, \textit{Phys.Rev.E} {\em{82}} 015304(R) (2010)
\bibitem{2} M.Sandoval, \textit{Phys.Rev.E} {\em{87}} 032708 (2013)
\bibitem{3} F.Kummel, B.ten Hagen, R.Witkowski, I.Buttinoni, G.Volpe, H.Lowen, and C.Bechniger, arXiv:1302.5787v (23 February 2013)
\bibitem{4} B.M.Friedrich, F.Julicher, \textit{New Journal of Physics} {\em{10}} 123025 (2008)
\bibitem{5} D.Takagi, A.B.Braunschweig, J.Zhang, and M.J.Shelley, \textit{Phys.Rev.Lett.} {\em{110}} 038301 (2013)
\bibitem{6}K.Erglis, Qi Wen, V.Ose, A.Zeltins, A.Sharipo, P.A.Janmey, and A.Cebers, \textit{Biophysical Journal} {\em{93}} 1402 (2007)
\bibitem{7} A.Cebers, \textit{JMMM} {\em{323}} 279 (2011)
\bibitem{8} K.Martens, L.Angelani, R.Di Leonardo, and L.Bocquet, \textit{Eur.Phys.J.E} {\em{35}} 84 (2012)
\bibitem{9} H.Berg, \textit{Random walks in Biology}, \textit{Princeton University Press}, (1983)
\bibitem{10} Ph.Nelson, \textit{Biological Physics}, \textit{W.H.Freeman and Company, New York} (2003)
\bibitem{11} C.Douarche, A.Buguin, H.Safman, and A.Libchaber, \textit{Phys.Rev.Lett.} {\em{102}} 198101 (2009)
\bibitem{12} E.Lauga, \textit{Phys.Rev.Lett.} {\em{106}} 178101 (2011)
\bibitem{13} I. Sendifia-Nadal, S.Alonso, V.Perez-Munuzuri, M.Gomez-Gesteira, V.Perez-Villar, L.Ramirez-Piscina, J.Casademunt, J.M.Sancho, and F.Sagues, \textit{Phys.Rev.Lett.} {\em{84}} 2734 (2000)
\end{thebibliography}
\end{document}